\documentstyle[aps,twocolumn]{revtex}

\begin{document}
\draft

\twocolumn[\hsize\textwidth\columnwidth\hsize\csname@twocolumnfalse\endcsname

\title{Simulation of many-body interactions   
by conditional geometric phases }
\author{ Xiaoguang Wang$^1$, Paolo Zanardi$^{1,2}$ }
\address{1. Institute for Scientific Interchange (ISI) Foundation, Viale Settimio Severo 65,
I-10133 Torino, Italy}
\address{2. Italy Istituto Nazionale di Fisica della Materia (INFM)}
\date{\today}
\maketitle

\begin{abstract}
It is shown  how to exactly simulate many-body interactions  
and multi-qubit  gates by coupling finite dimensional
systems, e.g., qubits with a continuous variable. 
Cyclic  evolution  in the phase space of such a variable gives rise
to  a geometric phase, 
depending  on a product of commuting operators.  
The latter   allows to 
simulate many-body Hamiltonians and nonlinear Hamiltonians, and 
to implement  a big variety 
of  multi-qubit quantum gates on both qubits and encoded qubits.
An application to the quantum amplitude amplification algorithm
will be discussed.
\end{abstract}

]

\section{Introduction}

\narrowtext
One of the most challenging and interesting problems in theoretical physics
concerns the understanding of quantum many-body systems. In view of the
exponentially many degrees of freedom involved, it is generally agreed that
simulation of such systems is a task that cannot be efficiently tackled by
any classical computers. On the other hand, as suggested by Feynman \cite
{Feynman} the growth in computational resources is only linear on a quantum
computer \cite{Qc}, which is itself a quantum many-body system. Therefore
even a modest size quantum computer, e.g., containing only a few tens of
qubits, could outperform a classical computer. We should like to stress that
the feasibility of such a ``devoted'' quantum computer problem is expected
to be definitely greater than the one of a general purpose quantum computer
\cite{Feynman,Lloyd1}.

Simulation of Hamiltonians has very practical applications in quantum control
\cite{Control} and quantum information theory \cite{QIP}. A typical goal is to achieve,
i.e., simulate, as many as possible desired entangling Hamiltonians starting
from a given one and the ability to perform local operations \cite
{Marcus,Dodd,Wocjan,Bennett1,Chen} or more generally to enact
transformations drawn from a given ``natural'' set of available interactions.

If we have two Hamiltonians $H_1$ and $H_2$ we can simulate the Hamiltonian $\lambda_1 H_1 +\lambda_2 H_2$ ($\lambda_1$ and $\lambda_2$ are real numbers) 
and the Hamiltonian $i[H_1,H_2]$ \cite{Deutsch}.  
Thus any Hamiltonians in the algebra generated by $H_1$ and $H_2$ can be simulated. This simulation method does not require that the two Hamitonians $H_1$ and $H_2$ commute but require that one can alternately turn on them for arbitrarily short time over infinite steps. Physically switching Hamiltonians over arbitrarily short time is not a easy task and unrealistic in some sense. We shall propose a method to do exact Hamiltonian simulations over finite steps.

In this paper we consider a ``hybrid'' physical system consisting of $%
N$ qubits coupled with a continuous variable system. The state space is then
given by ${\cal H}=({\rm \kern.24em\vrule width.04emheight1.46exdepth-.07ex%
\kern-.30emC}^2)^{\otimes \,N}\otimes h_\infty ,$ where $h_\infty :=%
\mbox{span}\{|n\rangle \}_{n=0}^\infty ,$ is the standard Fock space for a
single mode described by the bosonic annihilation and creation operators $a$ and 
$a^{\dagger }$, respectively. We assume the following Hamiltonians \cite{Monroe96,Gerry} to
be realizable 
\begin{eqnarray}
H_1 &=&-i\lambda (\alpha a^{\dagger }-\alpha ^{*}a)\text{ }\hat{A}
\label{eq:h1} \\
H_2 &=&\omega a^{\dagger }a\text{ }\hat{A},\text{ }
\end{eqnarray}
The operator $\hat{A}$ is a pure qubit operator that in the sequel will be
mostly either a Pauli operator $\sigma _{i\alpha }\,(\alpha =x,y,z)$ for ion 
$i$ or a collective angular momentum operator $J_\alpha =\frac 12%
\sum_{i=1}^N\sigma _{i\alpha }.$ The case of $\hat{A}=J_\alpha $ corresponds
to $N$ ions with each driven by identical Raman lasers. From the
Hamiltonians above we can have the conditional displacement operator and the
conditional rotation operator 
\begin{eqnarray}
U_1 &=&e^{-iH_1t}=e^{-\lambda t\hat{A}(\alpha a^{\dagger }-\alpha ^{*}a)%
\text{ }}, \\
U_2 &=&e^{-iH_2t}=e^{-i\omega t\hat{A}a^{\dagger }a\text{ }},
\end{eqnarray}
respectively. These two operators will play a crucial role in this paper in
which we will try to address the following question: given these two
kinds of physical operators, what is the maximal set of Hamiltonians that we
can simulate? 
In Ref. \cite{Milburn1} it has been shown that a conditional displacement of
the vibrational mode of trapped ions can be used to simulate nonlinear
collective and interacting spin systems including nonlinear tops and a
universal two--qubit gate, independent of the vibrational state of the ion.
The scheme in \cite{Milburn1} has been further extended in order to realize
the nonlinear Hamiltonian $J_z^{2\text{ }}$\cite{Anders} and multi-qubit
quantum gates \cite{Wangate}.

A general framework for quantum information processing (QIP) with hybrid
quantum systems has been proposed in Ref. \cite{Hybrid}. There it has been
proved that an universal set of Hamiltonians for hybrid quantum computation
is provided by the following: $\{\pm \sigma _x\hat{x},\pm \sigma _z\hat{x},
\pm \sigma _z\hat{p}\}.$ The position and momentum operators $\hat{x}$ and $%
\hat{p}$ satisfy the canonical commutation relation $[\hat{x},$ $\hat{p}]=i$%
. The ability to turn on and off the Hamiltonians from this set allows one
to produce conditional displacements that in turns allow to enact
Hamiltonians that are arbitrary polynomials of the $\sigma _x,$ $\sigma
_y,\sigma _z,\hat{x},$ and $\hat{p}.$ We can use an alternative set of
Hamiltonians $\{\sigma _xa^{\dagger }a,\sigma _za^{\dagger }a,\hat{x}\}.$
This set can be easily reduced to the above one as follows: 
\begin{eqnarray}
\pm \sigma _x\hat{x} &=&e^{\mp \frac \pi 2\sigma _xa^{\dagger }a}\hat{p}%
\text{ }e^{\pm \frac \pi 2\sigma _xa^{\dagger }a} \\
\pm \sigma _z\hat{x} &=&e^{\mp \frac \pi 2\sigma _za^{\dagger }a}\hat{p}%
\text{ }e^{\pm \frac \pi 2\sigma _za^{\dagger }a} \\
\pm \sigma _z\hat{p} &=&e^{\pm \frac \pi 2\sigma _za^{\dagger }a}\hat{x}%
\,e^{\mp \frac \pi 2\sigma _za^{\dagger }a}  \label{eq:333}
\end{eqnarray}
The operator $\hat{p}$ can be obtained from (\ref{eq:333}) when the spin is
in the state $|0\rangle $ or $|1\rangle$. These equations show that not only
conditional displacements but also conditional rotations are useful in
quantum computers. We will use both the conditional displacement operators
and conditional rotation operators to simulate many--body interaction
Hamiltonians and implement various quantum gates \cite{Gates,Wangate} and
algorithms \cite{Shor,Grover}.

Our scheme is interestingly related to one kind of geometric phase, i.e, the
geometric phase in phase space \cite{Phase,Luis}. However now our geometric
phase is dependent on some operators, i.e., it is a conditional geometric
phase. Recently both the adiabatic \cite{Berry} and non-adiabatic \cite{AA}
geometric phases have been suggested as potential candidates for realizing
quantum computers displaying some built-in fault tolerant features \cite
{Paolo,Jones,Fazio,Luming,Xiang}. We shall see that even geometric phases in phase
space could be useful for QIP.

The paper is organized as follows. In Sec. II we first review the concept of
geometric phase in phase space. In Sec. III we show how to generalize this
phase to `conditional' geometric phase and simulate many--body interactions.
Once these Hamiltonians are available, the implementation of some quantum
gates become somehow straightforward. We discuss the implementation of
quantum gates on both qubits and encoded qubits in Sec. IV. In Sec. V we
first show how to realize general projectors and then how to implement
the quantum amplitude amplification algorithm using the projectors. 
The conclusions are given in Sec. VI.

\section{ Geometric phases in phase space}

We begin by a brief review of the geometric phase within the formalism
recently given by Lius \cite{Luis}. Phase-space translations are represented
by the displacement operator 
\begin{equation}
{\cal D}\left( \alpha \right) \equiv D\left( \sqrt{2}\alpha \right) =e^{i(p%
\hat{x}-x\hat{p})},
\end{equation}
where the complex quantity $\alpha =x+ip$ parametrizes the displacement and $%
D\left( \alpha \right) =\exp (\alpha a^{\dagger }-\alpha ^{*}a)$ is the
usual displacement operator in quantum optics. These operators satisfy the
relation

\begin{equation}
{\cal D}(\beta ) {\cal D}(\alpha )= e^{\frac{i}{2}\text{Im}(\beta \alpha
^{*})} {\cal D} (\alpha+\beta ),  \label{eq:ddd}
\end{equation}
If $\alpha =x_1+ip_1$ and $\beta =x_2+ip_2$ one has $\text{Im}(\beta \alpha
^{*})=x_1 p_2-x_2 p_1$.

Eq.(\ref{eq:ddd}) tells us that the displacement operators realize an
unitary (projective) representation of the additive group of complex
numbers. In particular one has ${\cal D}(0)=1$, and ${\cal D}(-\alpha
)={\cal D}(\alpha )^{\dagger }.$ From Eq (\ref{eq:ddd}) it follows also the
useful identity

\begin{equation}
e^{i\text{Im}(\beta \alpha ^{*})}={\cal D}(-\beta ){\cal D}(-\alpha ){\cal D}%
(\beta ){\cal D}(\alpha ).  \label{eq:dd}
\end{equation}

We consider now a closed loop $\gamma $ which is an $N$--sided polygon with
sides $\alpha _j$ $(j=1,2,...,N) $ such that $\sum_{j=1}^N\alpha _j=0.$ The
total transformation associated with $\gamma $ is given by \cite{Luis}

\begin{equation}
{\cal D}_\gamma ={\cal D}(\alpha _N)\cdot \cdot \cdot {\cal D}(\alpha _1).
\label{eq:gamma}
\end{equation}

An arbitrary closed loop $\gamma$ can be approached in the limit of $%
N\rightarrow \infty $. In this case the total transformation is just a phase
factor ${\cal D}_\gamma =\exp (i\Theta )$, where

\begin{equation}
\Theta =\frac 12\oint (xdp-pdx).
\end{equation}
The phase $\Theta $ neither depends on the form of the loop nor on the speed
of the transformation but just on the area of the loop. For this reason it
deserves the name of a geometric phase.

\section{Simulation of many--body interactions}

Let us recall the Hamiltonian simulation \cite{Bennett1}. There is a set of
Hamiltonians $\{H_i\}$ and a class of allowed operations like local
unitaries or local operations and classical communications. The aim is to
produce a desired evolution $e^{-iHt}$ from the set of the Hamiltonians and
the class of allowed operations, where $H$ is the simulated Hamiltonian the $%
t$ is the simulated time.

Sometimes the class of local operations is not cheap resource. In this
paper we will not use them and just begin from a set of Hamiltonians to do
the simulation.

\subsection{Nonlinear Hamiltonians and two--body interactions}

A conditional displacement operator reads
${\cal D}(\hat{A}\alpha)$,
where the Hermitian operator $\hat{A}$ usually represents a discrete
variable observable. 
By using the spectrum decomposition of the operator 
$\hat{A}$, $\hat{A} 
=\sum_{k=1}^N
\lambda _k
|k\rangle \langle k|$ the
displacement operator can be written as

\begin{equation}
{\cal D}(\hat{A}\alpha )=\sum_{k=1}^N|k\rangle \langle k|{\cal D}(\lambda
_k\alpha ).
\end{equation}
From the above equation one can see clearly that the amount of displacement is
dependent on the eigenvalues of $\hat{A}.$ By replacing the displacement
operators in Eq.(\ref{eq:dd}) with conditional displacement operators, we
immediately obtain

\begin{equation}
e^{i\theta \hat{A}\hat{B}}={\cal D}(-\hat{B}\beta ){\cal D}(-\hat{A}\alpha )%
{\cal D}(\hat{B}\beta ){\cal D}(\hat{A}\alpha )\text{ ,}  \label{eq:ab}
\end{equation}
where $\theta =$Im$(\beta \alpha ^{*})$ and operator $\hat{B}$ commutes with 
$\hat{A}.$ Now the geometric phase depends on the operator product $\hat{A}%
\hat{B}.$ Therefore we can simulate the Hamiltonian $\hat{A}\hat{B}$ if
operators $\hat{A}$ and $\hat{B}$ commute and belong to one system. For
example if $\hat{A}$ $=\hat{B}=J_z,$ we get the nonlinear Hamiltonian $J_z^2,
$ which is useful for creating GHZ multipartite entangled states \cite
{GHZ,Molmer} and generate spin--squeezed states \cite{Ueda}. If $\hat{A}$ and 
$\hat{B}$ belong to different systems we may simulate the two--body
interaction Hamiltonian $\hat{A}\otimes \hat{B}$.

\subsection{Three--body interactions}

In the context of NMR quantum computation, Tseng {\it et al.}
\cite{Three} proposed
a method to simulate three--body interaction of the type $\sigma _z\otimes
\sigma _z\otimes \sigma _z.$ Due to the self--inverse property of $\sigma _z$
simulation of the three--body Hamiltonian becomes relatively easy by
two--body interactions. The simulation can be obtained in the following way

\begin{eqnarray}
e^{-i\theta \sigma _{1z}\otimes \sigma _{2z}\otimes \sigma _{3z}} &=&e^{-i%
\frac \pi 4\sigma _{1x}\otimes \sigma _{3z}}e^{-i\frac \pi 4\sigma
_{1x}\otimes \sigma _{2z}}e^{i\theta \sigma _{1z}}  \nonumber \\
&&e^{i\frac \pi 4\sigma _{1x}\otimes \sigma _{3z}}e^{i\frac \pi 4\sigma
_{1x}\otimes \sigma _{2z}}.
\end{eqnarray}
Now we show how to create `nonphysical' three--body interaction Hamiltonians 
$H=\hat{A}\otimes \hat{B}\otimes \hat{C}$ for general operators $\hat{A},%
\hat{B},$ and $\hat{C}$ which act on three different subsystems,
respectively. To this aim let us introduce the operator ${\cal R}(\theta
)=\exp (i\theta a^{\dagger }a),$ it entails a rotation in the phase space

\begin{equation}
{\cal R}(-\theta )\left( 
\begin{array}{l}
\hat{x} \\ 
\hat{p}
\end{array}
\right) {\cal R}(\theta )=\left( 
\begin{array}{ll}
\cos \theta & -\sin \theta \\ 
\sin \theta & \cos \theta
\end{array}
\right) \left( 
\begin{array}{l}
\hat{x} \\ 
\hat{p}
\end{array}
\right) .  \label{eq:xp}
\end{equation}
Directly from the above equation we obtain

\begin{equation}
{\cal D}(\alpha e^{i\theta })={\cal R}(\theta ){\cal D}(\alpha ){\cal R}%
(-\theta ).  \label{eq:r}
\end{equation}
The conditional rotation operator is given by ${\cal R}(\theta \hat{C}),$
which makes a rotation in phase space conditioned on the operator $\hat{C}.$
As in Eq.(\ref{eq:r}) we get

\begin{equation}
{\cal D}(\alpha e^{i\theta \hat{C}})={\cal R}(\theta \hat{C}){\cal D}(\alpha
){\cal R}(-\theta \hat{C}).
\end{equation}
Then we can replace $\alpha $ with $\alpha e^{i\theta \hat{C}}$ in Eq.(\ref
{eq:ab}) and obtain

\begin{eqnarray}
&&e^{-i\tau \hat{A}\otimes \hat{B}\otimes \sin (\theta \hat{C}+\phi )} 
\nonumber \\
&=&{\cal D}(-\hat{B}\beta ){\cal D}(-\hat{A}\alpha e^{i\theta \hat{C}}){\cal %
D}(\hat{B}\beta ){\cal D}(\hat{A}\alpha e^{i\theta \hat{C}}),  \label{eq:abc}
\end{eqnarray}
where $\tau =|\alpha \beta |$ and $\phi =\arg (\alpha )-\arg (\beta ).\,$
Eq.(\ref{eq:abc}) tells us that we can simulate the three--body Hamiltonian 
\begin{equation}
H=\lambda \hat{A}\otimes \hat{B}\otimes \sin (\theta \hat{C}+\phi )
\label{eq:aaa}
\end{equation}
by the following sequence

\begin{eqnarray}
&&{\cal D}(-\hat{B}\beta ){\cal R}(\theta \hat{C}){\cal D}(-\hat{A}\alpha )%
{\cal R}(-\theta \hat{C})  \nonumber \\
&&\times {\cal D}(\hat{B}\beta ){\cal R}(\theta \hat{C}){\cal D}(\hat{A}%
\alpha ){\cal R}(-\theta \hat{C}).  \label{eq:abab}
\end{eqnarray}
Therefore the `nonphysical' three--body interaction is {\em exactly} achieved by
eight two--body physical interactions.

Specifically we obtain the Hamiltonian

\begin{equation}
H=\lambda \hat{A}\otimes \hat{B}\otimes \sin (\theta \hat{C}).
\label{eq:absinc}
\end{equation}
If we make a small rotation, i.e., the angle $\theta $ is small enough, we
can approximately realize the three-body interaction Hamiltonian

\begin{equation}
H=\lambda^\prime\hat{A}\otimes \hat{B}\otimes \hat{C}.
\label{eq:threethree}
\end{equation}
among three operators $\hat{A}$, $\hat{B}$ and $\hat{C}$, where $\lambda
^\prime =\lambda \theta .$ Remarkably for some operator $\hat{C}$ we
can even exactly realize the three--body interaction.

Let us consider an operator $\hat{C}$ which satisfies either $\hat{C}^2=1$
(self-inverse operators)
or $\hat{C}^2=\hat{C}$ (idempotent operators). 
For such an operator we have $\sin (\theta \hat{C}%
)=\sin \theta \hat{C}.$ Therefore Eq.(\ref{eq:absinc}) shows that we can
exactly realize the Hamiltonian $\hat{A}\otimes \hat{B}\otimes \hat{C}.$ The
self--inverse operators we often encounter are Pauli operators and parity
operators {\it et al}. Most quantum gates in quantum computer are also
self--inverse, such as Hadmard gate, controlled--NOT gate ({\bf C}$_{\text{NOT}}$)\cite{Cnot}, 
SWAP gate and Toffoli gate\cite{Gates}. Examples of idempotent operators are of
course given by projection operators.

Summarizing the three--body interaction Hamiltonian $\hat{A}\otimes \hat{B}%
\otimes \hat{C}$ is approximately simulated for any operators $\hat{A},%
\hat{B},$ and $\hat{C},$ and is exactly simulated if one of the operators is
self--inverse or idempotent. For the case of $\hat{A}=\hat{B}=\hat{C}=J_z,$
we approximately have the Hamiltonian $J_z\otimes J_z\otimes J_z.$ By
choosing the self--inverse $\hat{C}=\sigma _z,$ we exactly have the
Hamiltonians $J_z\otimes J_z\otimes \sigma _z$, and $\sigma _z\otimes \sigma
_z\otimes \sigma _z.$

Both Eq.(\ref{eq:ab}) and Eq.(\ref{eq:abab}) correspond to a close loop
(a parallelogram) in phase space. Now the resulting geometric phases are
dependent on a product of commuting operators, which are just the desired
simulated Hamiltonians.

\subsection{Many--body interactions of qubits}

In this subsection we address the simulation problem of some many--body
interactions of qubits which are related to quantum spin models in condensed
matter theory. From Eq.(\ref{eq:aaa}) we have the Hamiltonian

\begin{eqnarray}
H_c &=&\lambda \cos (\theta J_\alpha ),  \label{eq:sincc} \\
H_s &=&\lambda \sin (\theta J_\alpha ).  \label{eq:sinc}
\end{eqnarray}
Let $\theta =\pi $ we obtain

\begin{eqnarray}
H_c &=&\lambda \cos (\pi N/2)\sigma _\alpha \otimes \sigma _\alpha \otimes
...\otimes \sigma _\alpha ,  \label{eq:coscos} \\
H_s &=&\lambda \sin (\pi N/2)\sigma _\alpha \otimes \sigma _\alpha \otimes
...\otimes \sigma _\alpha .  \label{eq:sinsin}
\end{eqnarray}
From Eq.(\ref{eq:coscos}) with even $N$ and Eq.(\ref{eq:sinsin}) with odd $%
N, $ we can simulate, for arbitrary $N,$ the multi--qubit Hamiltonians
\begin{equation}
H_{\alpha \pm }=\pm \lambda \sigma _\alpha \otimes \sigma _\alpha \otimes
...\otimes \sigma _\alpha ,  \label{eq:hhh}
\end{equation}
The above many--body Hamiltonians are very useful. For instance, let the
initial state the $N$--qubit system be $|0\rangle ^{\otimes N}$ and evolve
according to the Hamiltonian $H_{x+},$ then the state vector at time $t\,$
is easily obtained as

\begin{equation}
|\theta ,N\rangle =\cos (\theta )|0\rangle ^{\otimes N}-i\sin (\theta
)|1\rangle ^{\otimes N},
\end{equation}
where $\theta =\lambda t.$ The state $|\pi /4,N\rangle $ is a multipartite
entangled state, which is known to play an important role in quantum
information theory.

From Eq.(\ref{eq:hhh})$,$ we have the Hamiltonian $\sigma _{1\alpha }\sigma
_{2\alpha }(\alpha =x,y,z)$ for $N=2$, which are just the Ising
interactions. As $\sigma _{1x}\sigma _{2x},\sigma _{1y}\sigma _{2y},$ and $%
\sigma _{1z}\sigma _{2z}$ commutes with each other, we may simulate a
general two--qubit Heisenberg model \cite{Heis1}

\begin{equation}
H=\lambda_x\sigma _{1x}\sigma _{2x}+\lambda_y\sigma _{1y}\sigma _{2y}
+\lambda_z\sigma
_{1z}\sigma _{2z}.  \label{eq:heis}
\end{equation}

For $N>2,$ i.e., the many--body case, we already have the Hamiltonians $%
\sigma _x^{\otimes N},$ $\sigma _y^{\otimes N},$ and $\sigma _z^{\otimes N},$
which satisfy the commutation relation 
\begin{equation}
\lbrack \sigma _x^{\otimes N},\sigma _y^{\otimes N}]_{\pm }=i^N[1\pm
(-1)^N]\sigma _z^{\otimes N}.  \label{eq:delta}
\end{equation}
Here the subscripts $+$ and $-\,$ indicates the anticommutation and
commutation, respectively. From the above equation we see that the three
operators either commute or anticommute. For even $N,$ any two of the three
operators commute with each other. Then we can simulate the following
Hamiltonian\cite{Heismulti}

\begin{eqnarray}
H &=&\lambda_x\sigma _x^{\otimes (2k)}+\lambda_y\sigma _y^{\otimes (2k)}
+\lambda_z\sigma
_z^{\otimes (2k)}\text{ }  \label{eq:kkk} \\
\text{(}k &=&1,2,3...\text{)}.  \nonumber
\end{eqnarray}

However for odd $N,$ any two of the three operators anticommute but do not
commute with each other. For $N=4m+1$ $(m=0,1,2,...),$ from Eq.(\ref
{eq:delta}), we obtain

\begin{equation}
\lbrack \sigma _x^{\otimes N},\sigma _y^{\otimes N}]_{-}=2i\sigma
_z^{\otimes N}.  \label{eq:encode}
\end{equation}
Thus the operators $\sigma _x^{\otimes N}/2,\sigma _y^{\otimes N}/2$ and $%
\sigma _z^{\otimes N}/2$ realize the su(2) Lie algebra. For odd $N=4m+3$ $%
(m=0,1,2,...),$ from Eq.(\ref{eq:delta}), we obtain

\begin{equation}
\lbrack \sigma _x^{\otimes N},\sigma _y^{\otimes N}]_{-}=-2i\sigma
_z^{\otimes N}.  \label{eq:code1}
\end{equation}
In this case $\sigma _x^{\otimes N}/2,\sigma _y^{\otimes N}/2$ and $-\sigma
_z^{\otimes N}/2$ realize the su(2) Lie algebra. For even $N$ we can find
three operators

\begin{equation}
\sigma _x^{\otimes (2k)}/2,\sigma _y\otimes \sigma _x^{\otimes
(2k-1)}/2,\sigma _z\otimes I^{\otimes (2k-1)}/2  \label{eq:code2}
\end{equation}
satisfy the su(2) commutation relations. These realizations of su(2) Lie
algebras will be used in the later discussions of quantum gates on encoded
qubits.

From any set of commuting Hamiltonians $\{\lambda _iH_i\}$ we can simulate
the Hamiltonian $\sum_i\lambda _iH_i$ since

\begin{equation}
\exp (\sum_i\lambda _i H_i)=\prod_i\exp (\lambda _iH_i).  \label{eq:decom}
\end{equation}
Even for noncommuting set $\{\lambda _iH_i\},$ we still have a chance that a
decomposition similar to Eq.(\ref{eq:decom}) exists\cite{KimLee}. For
instance, the operators in Eq.(\ref{eq:encode})$,\tilde{\sigma}_\alpha
=\sigma _\alpha ^{\otimes N},$ act as the encoded Pauli matrices$.$ Then we
have the identity

\begin{eqnarray}
&&\exp (-i\phi [\cos \theta \tilde{\sigma}_z+\sin \theta \tilde{\sigma}_y)/2]
\nonumber \\
&=&\exp (i\theta \tilde{\sigma}_x/2)\exp (-i\phi \tilde{\sigma}_z/2)\exp
(-i\theta \tilde{\sigma}_x/2),
\end{eqnarray}
which implies that we can simulate the Hamiltonian

\begin{equation}
H=\lambda _z\tilde{\sigma}_z+\lambda _y\tilde{\sigma}_y=\lambda _z\sigma
_z^{\otimes N}+\lambda _y\sigma _y^{\otimes N}.
\end{equation}
for odd $N=4m+1$ using $\tilde{\sigma}_x=\sigma _x^{\otimes N}$ and $\tilde{%
\sigma}_z=\sigma _z^{\otimes N}.$

\section{Simulation of quantum gates}

So far we have showed how to obtain, by cyclic conditional evolutions in
phase space, an operator--dependent geometric phase and how to use these
operators in order to simulate two--body and many--body interaction
Hamiltonians. Now we make use of the above-developed formalism to explicitly
construct some important quantum logic gates. We emphasize that the
continuous e.g., vibrational degree of freedom is only required during
gating and acts like a databus.

\subsection{Two-qubit gates}

Controlled--NOT gate\cite{Cnot} and controlled phase gate 
({\bf C}$_{\text{P}}$)\cite{CP}: Let $\hat{A}%
=(1-\sigma _{1z})/2$, $\hat{B}=(1-\sigma _{1x})/2,$ and $\theta =\pi $ in
Eq.(\ref{eq:ab}), the controlled--NOT gate is immediately obtained as

\begin{equation}
{\bf C}_{\text{NOT}}=\exp \left[ -i{\frac \pi 4}(1-\sigma _{1z})(1-\sigma
_{2x})\right] .
\end{equation}
The first bit is the control bit and the second is the target bit.
Similarly the controlled--phase gate is obtained as

\begin{equation}
{\bf C}_{\text{P}}=\exp \left[ -i\frac{{\pi }}4(1-\sigma _{1z})(1-\sigma
_{2z})\right].
\end{equation}
A simple relation exists between the controlled--NOT gate and the
controlled--phase gate,

\begin{equation}
{\bf C}_{\text{NOT}}=\exp \left[ -i\frac \pi 4\sigma _{2y}\right] {\bf C}_{%
\text{P}}\exp \left[ i\frac \pi 4\sigma _{2y}\right] ,
\end{equation}
i.e., they differ only by local operations.

SWAP gate : As we can simulate the Heisenberg
Hamiltonian $($\ref{eq:heis}), the SWAP gate is easily constructed as

\begin{eqnarray}
{\bf G}_{\text{SWAP}} &=&e^{-i\frac \pi 4\left( \sigma _{1x}\sigma
_{2x}+\sigma _{1y}\sigma _{2y}+\sigma _{1z}\sigma _{2z}-1\right) }  \nonumber
\\
&=&\frac 12\left( 1+\sigma _{1x}\sigma _{2x}+\sigma _{1y}\sigma _{2y}+\sigma
_{1z}\sigma _{1z}\right) .
\end{eqnarray}

\subsection{Three-qubit gates}

Toffoli gate ({\bf T})\cite{Gates}: Let $\hat{A}=(1-\sigma _{1z})/2,\hat{B}%
=(1-\sigma _{2z})/2,$ and $\hat{C}=$ $(1-\sigma _{3x})/2,$ in Eq.(\ref
{eq:threethree}), the three-bit Toffoli gate is obtained as

\begin{equation}
{\bf T}=\exp \left[ -i{\frac \pi 8}(1-\sigma _{1z})(1-\sigma _{2z})(1-\sigma
_{3x})\right] ,
\end{equation}
which is also called (controlled)$^2$-NOT (${\bf C}_{\text{NOT}}^2)$ gate
and important for universal computation. Note the Toffoli gate is exactly
realized since the operator $\hat{C}=$ $(1-\sigma _{3x})/2$ is a projector.
Such a gate can of course be also implemented by constructing appropriate
networks of one qubit and two qubit gates \cite{Gates}. The construction
showed here is somehow more direct. Notice also that an alternative way to
obtain the Toffoli gate is recently discussed in Ref. \cite{Wangate}. The
implementation of the more general (controlled)$^N$-NOT will be discussed in
Section V.

Fredkin gate ({\bf F}): Another example of a three qubit gate relevant for
QIP is provided by the controlled--SWAP Fredkin gate\cite{Fredkin}. The
following {\em commuting Hamiltonians } $(1-\sigma _{1z})\,\otimes \sigma
_{2\alpha }\otimes \sigma _{3\alpha }(\alpha =x,y,z)$ can be realized
exactly (\ref{eq:absinc}). Therefore we can simulate the unitary operator
\begin{equation}
\text{{\bf F}}=e^{-i\frac \pi 8(1-\sigma _{1z})\left( \sigma _{2x}\sigma
_{3x}+\sigma _{2y}\sigma _{3y}+\sigma _{2z}\sigma _{3z}-1\right) }, 
\end{equation}
which is just the Fredkin gate.

\subsection{Quantum gates on encoded qubits}

In order to perform universal quantum computations it is sufficient to be
able to make arbitrary single qubit rotations together with
controlled--phase gate. For encoded qubits one problem is how to make
logical operations \cite{Zurek} of them and the above two logical operations on
encoded qubits are needed. We will discuss two typical codes: the active
error correction codes \cite{Steane} and the passive codes on decoherence-free subspaces 
\cite{DFS}.

First we consider the error correction codes with odd number, the linear
codes proposed by Steane \cite{Steane}. He have devised two encoding, the
first of which protects only against decoherence

\begin{eqnarray}
|0_{\text{C}}\rangle &=&\frac 12\left( |111\rangle +|100\rangle +|010\rangle
+|001\rangle \right) ,  \nonumber \\
|1_{\text{C}}\rangle &=&\sigma _x\otimes \sigma _x\otimes \sigma _x|0_{\text{%
C}}\rangle ,  \label{eq:threebit}
\end{eqnarray}
The second is capable of decoding with general 1-bit errors

\begin{eqnarray}
|0_{\text{C}}\rangle &=&\frac 1{\sqrt{8}}(|1111111\rangle +|0101010\rangle
+|1001100\rangle  \nonumber \\
&&+|0011001\rangle +|1110000\rangle +|0100101\rangle  \nonumber \\
&&+|1000011\rangle +|0010110\rangle ),  \nonumber \\
|1_{\text{C}}\rangle &=&\sigma _x\otimes \sigma _x\otimes \sigma _x\otimes
\sigma _x\otimes \sigma _x\otimes \sigma _x\otimes \sigma _x\,|0_{\text{C}%
}\rangle .
\end{eqnarray}

The encoded qubits span a representation space of su(2) Lie algebra
generated by the operators $\sigma _x^{\otimes N},\sigma _y^{\otimes N}$ and
--$\sigma _z^{\otimes N}$ (\ref{eq:code1})$.$ Here $N$ is either 3 or 7.
That is to say, these three operators acts on encoded qubits as encoded $%
\sigma _x,\sigma _y$ and $\sigma _z.$ Then we consider the code mapping 1 qubit
into 5 qubits presented in Ref. \cite{Bennett}

\begin{eqnarray}
|0_{\text{C}}\rangle &=&Q|00000\rangle ,\text{ }|1_{\text{C}}\rangle
=Q|11111\rangle ,  \nonumber \\
Q &=&\frac 14[\openone ^{\otimes 5}+(\sigma _x\otimes \sigma _x\otimes %
\openone \otimes \openone \otimes \openone )_{cyc}  \nonumber \\
&&-(\sigma _x\otimes \openone \otimes \sigma _x\otimes \openone \otimes %
\openone )_{cyc}  \nonumber \\
&&-(\sigma _x\otimes \sigma _x\otimes \sigma _x\otimes \sigma _x\otimes %
\openone )_{cyc}],
\end{eqnarray}
where the subscript ``cyc'' indicates that addition of all five cyclic
shifts. Obviously the commutators [$\sigma _\alpha ^{\otimes 5},Q]=0$ ($%
\alpha =x,y,z$) and then $\sigma _x^{\otimes 5}$ $\sigma _y^{\otimes 5},$
and $\sigma _z^{\otimes 5}$(\ref{eq:encode}) act as encoded $\sigma _x$, $%
\sigma _y$ and $\sigma _z,$ respectively $.$

Now we consider even--number codes in decoherence--free subspace \cite{DFS}.
We can have a code of two qubits \cite{Hwang}

\begin{equation}
|0_{\text{C}}\rangle =|01\rangle ,\text{ }|1_{\text{C}}\rangle =|10\rangle .
\end{equation}
and a code of four qubits

\begin{eqnarray}
|0_{\text{C}}\rangle &=&|0011\rangle +|0110\rangle +|0101\rangle , \\
|1_{\text{C}}\rangle &=&|1100\rangle +|1001\rangle +|1010\rangle ,  \nonumber
\end{eqnarray}
For these two cases, it is easy to see that the two-dimensional subspace span$%
\left\{ |0_{\text{C}}\rangle ,|1_{\text{C}}\rangle \right\} $ is
decoherence-free subspace. The corresponding encoded Pauli matrices are given by
Eq.(\ref{eq:code2}).

The controlled--phase gate for encoded qubits is easily constructed as 
\begin{equation}
{\bf \tilde{C}}_{\text{P}}=\exp \left[ -i\pi \left( \frac{1-\tilde{%
\sigma}_z}2\right) \otimes \left( \frac{1 -\tilde{\sigma}_z}2\right)
\right] .
\end{equation}
Explicitly for three--bit code we give the gate as

\begin{eqnarray}
{\bf \tilde{C}}_{\text{P}} &=&\exp [-i\frac \pi 4(\openone ^{\otimes 6}+%
\openone ^{\otimes 3}\otimes \sigma _z^{\otimes 3}  \nonumber \\
&&+\sigma _z^{\otimes 3}\otimes \openone ^{\otimes 3}+\sigma _z^{\otimes 6})]
\end{eqnarray}
which can be realized by our simulation method. Then we have both the
rotations of single encoded qubit and gate ${\bf \tilde{C}}_{\text{P}}$ for
two encoded qubits, which are enough for the quantum computation on the
encoded qubits.

\section{Applications to quantum algorithms and discussions}

In this section we show some applications of the simulation schemes
discussed so far to the implementation of quantum algorithms.

\subsection{Realization of projectors}

Let us consider the Hamiltonian 
\begin{equation}
H_k=\lambda \cos [\theta _k(J_z+\phi _k)].
\end{equation}
Since the Hamiltonians $H_k\,$commute with each other for two different $k,$ we
have the sum of them

\begin{equation}
H=\lambda \sum_{k=0}^N\cos \left[ \theta _k(J_z+\phi _k)\right]
\end{equation}
This Hamiltonian can be obtained by $N+1$ cyclic evolutions in phase space.
If we choose $\theta _k=\frac{2\pi k}{N+1}$ $,\lambda =\frac 1{N+1}$and $%
\phi _k=N/2+n$ ($k=0...N$), the Hamiltonian becomes

\begin{equation}
H=\frac 1{N+1}\sum_{k=0}^N\cos \left[ \frac{2\pi k}{N+1}\left( {\cal N}%
-n\right) \right] ,
\end{equation}
where ${\cal N}=J_z+N/2$ whose eigenvalues range from $0$ to $N.$ From the
above equation and using the identity

\begin{equation}
\frac 1{N+1}\sum_{k=0}^N\cos \left[ \frac{2\pi k}{N+1}\left( n-n^{\prime
}\right) \right] =\delta _{nn^{\prime }}\text{,}
\end{equation}
we obtain the Hamiltonian

\begin{equation}
H=\delta _{{\cal N}n} 
\end{equation}
which implies that we have realized the projector $P_n$ which project the
state to the symmetric subspace with excitation $n.$ For instance, for $n=0$
or $N,$ the projectors are just

\begin{eqnarray}
P_0 &=&|00...0\rangle \langle 00...0|, \\
P_N &=&|11...1\rangle \langle 11...1|.
\end{eqnarray}
A general Hamiltonian $F(J_z)$ can be written as

\begin{equation}
F(J_z)=\sum_{n=0}^NF(n-N/2)P_n.
\end{equation}
Since all $P_n$ commute we can simulate the general Hamiltonian $F(J_z).$

Once we have realized the projector $P_0$ we can realize any one of $2^N$
projectors

\begin{equation}
{\cal P}_{a_1a_2...a_n}=|a_1,a_2,...,a_n\rangle \langle a_1,a_2,...,a_n|
\label{eq:ppp}
\end{equation}
in the $N$--qubit space, where $a_i\in \{0,1...N\}$ and the state $%
|a_1,a_2,...,a_n\rangle $ represents that the only in the positions of $%
a_1,a_2,$ and $a_n$ the qubit is in the state $|1\rangle $ and in the state $%
|0\rangle $ in other positions. The projector can be implemented as follows

\begin{equation}
{\cal P}_{a_1a_2...a_n}=\sigma _{a_1x}\sigma _{a_2x},...,\sigma
_{a_Nx}P_0\sigma _{a_1x}\sigma _{a_2x},...,\sigma _{a_Nx}.
\end{equation}
Then we realize all the projectors in the $N$--qubit space. This can be
understood from a more general point of view \cite{Dodd}. Suppose we have the
Hamiltonian $H$ and perform unitary operations $U$ and $U^{\dagger }.$ Then
it follows from the identity $e^{-itUHU^{\dagger }}=Ue^{-itH}U^{\dagger }$
that we can exactly simulate evolution according to the Hamiltonian $%
UHU^{\dagger }.$ Now our $H$ and $U$ are: $H=P_0$ and $U=U^{\dagger }=\sigma
_{a_1x}\sigma _{a_2x},...,\sigma _{a_Nx}.$

The projector ${\cal P}_{a_1a_2...a_N}$ is very useful. For instance, from $%
P_N:=|1\rangle \langle 1|^{\otimes \,N},$ the ($N+1)$--bit ${\bf C}_{\text{%
NOT}}^N$ gate, is expressed as\cite{Gates}

\begin{eqnarray}
{\bf C}_{\text{NOT}}^N &=&\exp \left[ -i{\frac \pi 2}\left( P_N\right)
(1-\sigma _{n+1x})\right]  \nonumber \\
&=&1-P_N+P_N\text{ }\sigma _{n+1x},
\end{eqnarray}
which is a natural generalization of the controlled-NOT gate and Toffoli
gate to many qubits. Alternatively one can construct the multi--qubit
generalization of the controlled--phase gate

\begin{equation}
U_{P_N}=e^{-i\pi P_N/2}
\end{equation}
Then the C$_{\text{NOT}}^{N-1}$ gate is easily obtained as

\begin{equation}
{\bf C}_{\text{NOT}}^{N-1}=e^{-i\frac \pi 4\sigma _{Ny}}U_{P_N}e^{i\frac \pi %
4\sigma _{Ny}}.
\end{equation}
So we can straightforwardly implement {\bf C}$_{\text{NOT}}^{N-1}${\bf \ }%
gate once we have had the projector.

\subsection{Implementation of the quantum amplitude amplification algorithm}

In 1997, Grover presented a search algorithm \cite{Grover} that identifies
the single value $x_0$ that fulfills $f(x_0)=1$ for a function $f(x)$
provided, {\it e.g.}, by an oracle (all other arguments lead to vanishing
values of the function). If $x$ is an integer on the range between 0 and $%
N-1=2^n-1$, the search algorithm is able to find $x_0$ after on the order of 
$\sqrt{N}$ evaluations of the function. Grover's algorithm has been
demonstrated on NMR few qubit systems \cite{NMRgrover}. In our previous
paper \cite{Wangate} we have shown how to implement the Grover's algorithm. In the
following we will use our method to implement a general quantum search,
which is called quantum amplitude amplification algorithm \cite{Hoyer}.

We can write a general quantum search operator \cite{Hoyer,Long} as

\begin{equation}
{\bf Q}={\bf Q(}{\cal A}{\bf ,\chi ,\varphi ,\vartheta )=-}{\cal A}{\bf S}%
_0^\varphi {\cal A}^{-1}{\bf S}_\chi ^\vartheta ,
\end{equation}
which is at the heart of the quantum algorithm. Here ${\cal A}$ is any
quantum algorithm that acts on the $N$--qubit system and

\begin{eqnarray}
{\bf S}_0^\varphi  &=&1+(e^{i{\bf \varphi }}-1)|00...0\rangle \langle
00...0|,  \label{eq:chi1} \\
{\bf S}_\chi ^\vartheta  &=&1+(e^{i{\bf \vartheta }}-1)\sum_k|{\bf \tau }%
_k\rangle \langle {\bf \tau }_k|.  \label{eq:chi2}
\end{eqnarray}
The state $|{\bf \tau }_k\rangle $ is a marked state and the summation runs
over all the marked states. Thus this quantum search is a multi--object
search. The quantum algorithm contains a unitary transformation, two phase
rotations and the marked states. When ${\cal A}$ is the Walsh-Hadmard
transformation, there is one marked state, and ${\bf \vartheta =\varphi =\pi
,}$ the quantum algorithm reduces to the usual Grover's search algorithm.

Now we write the two phase rotation operators ${\bf S}_0^\varphi $ and ${\bf %
S}_\chi ^\vartheta $ as exponential form as

\begin{eqnarray}
{\bf S}_0^\varphi &=&e^{i{\bf \varphi }|00...0\rangle \langle 00...0|}, \\
{\bf S}_\chi ^\vartheta &=&e^{i{\bf \vartheta }\sum_k|{\bf \tau }_k\rangle
\langle {\bf \tau }_k|}.
\end{eqnarray}
As we have all the projectors (\ref{eq:ppp}) and they commute with each
other, the two rotation operators are then realized straightforwardly.
Therefore the general quantum search algorithm can be implemented. In a
recent paper \cite{Lidafa} the phase rotation operator ${\bf S}_\tau
^\vartheta =1-2\cos \theta e^{i\vartheta }|{\bf \tau }\rangle \langle {\bf %
\tau }|$ is introduced. This operator can be written as the exponential form
as ${\bf S}_\tau ^\vartheta =e^{i(\pi -{\bf 2\vartheta )}|{\bf \tau }\rangle
\langle {\bf \tau }|}$ and therefore we can realize it similarly as ${\bf S}%
_0^\varphi $ and ${\bf S}_\chi ^\vartheta $.

A few comments on the above geometric scheme in this paper are now in order. From Eq.(\ref{eq:abc})
it can be seen that we still have freedom to make a generalization as

\begin{eqnarray}
&&e^{-i\tau _{jk}\hat{A}_j\hat{A}_k\sin (\theta _j\hat{C}_j-\theta _k\hat{C}%
_k+\phi _{jk})}={\cal D}(\hat{A}_k\alpha _ke^{i\theta _k\hat{C}_k}) 
\nonumber \\
&&\times {\cal D}(-\hat{A}_j\alpha _je^{i\theta _j\hat{C}_j}){\cal D}(\hat{A}%
_k\alpha _ke^{i\theta _k\hat{C}_k}){\cal D}(\hat{A}_j\alpha _je^{i\theta _j%
\hat{C}_j}).
\end{eqnarray}
Here $\tau _{jk}=|\alpha _j\alpha _k|$ and $\phi _{jk}=\arg (\alpha _j)-\arg
(\alpha _k).$ So we can simulate the Hamiltonian like $\hat{A}_j\hat{A}%
_k\sin (\theta _j\hat{C}_j-\theta _k\hat{C}_k+\phi _{jk})$ which includes
for arbitrary commuting operators $\hat{A}_j,$ $\hat{A}_k,\hat{C}_j,$ and $%
\hat{C}_k.$ Then we further ask if we can exactly simulate the product of
three arbitrary operators $\hat{A}_j\hat{A}_k\hat{C}_l\,$or more
general one, the product of four operators $\hat{A}_j\hat{A}_k\hat{C}_l\hat{C%
}_m.$ The answer seems negative with our scheme. The reason is as follows.
The general transformation is given by Eq.(\ref{eq:gamma}) with $\alpha
_i\rightarrow \alpha _iA_ie^{i\theta _i\hat{C}_i}.$ Then if the
transformation $\gamma $ is cyclic, we obtain a geometric phase factor given
by

\begin{equation}
e^{-i\sum_{j>k}\tau _{jk}\hat{A}_j\hat{A}_k\sin (\theta _j\hat{C}_j-\theta _k%
\hat{C}_k+\phi _{jk})}
\end{equation}
Hence we can not exactly achieve the products of three or more commuting
operators.

\section{ Conclusions}

In this paper we have discussed a geometric scheme to simulate many--body
interactions and to implement multi-qubit gates. Our strategy is
based on conditional cyclic evolutions in the phase space of a continuous
quantum variable coupled to discrete systems. The cyclic evolution leads to
a conditional geometric phase factor containing the desired operators acting
non trivially on the discrete factor. To the use of the reader let us list
the main results of this work:

1. We can {\em exactly} simulate two--body Hamiltonians like $\hat{A}\otimes 
\hat{B},$ the three--body interaction Hamiltonians like $\hat{A}%
\otimes \hat{B}\otimes \sin (\theta \hat{C}+\phi )$, $\hat{A}\otimes \hat{%
B}\otimes \hat{C}$ with one of them is self--inverse or idempotent, and
many--body Hamiltonians like $\sigma _\alpha \otimes \sigma _\alpha \otimes
...\otimes \sigma _\alpha .$ Some quantum spin models in condensed matter
theory are exactly simulated. We approximately simulated the Hamiltonian $%
\hat{A}\otimes \hat{B}\otimes \hat{C}$ for three operators $\hat{A}%
,\hat{B},$ and $\hat{C}.$ 2. We can simulate the nonlinear Hamiltonian $J_z^2
$ and the more general nonlinear Hamiltonian $F(J_z).$ 3. Nearly all the
quantum gates proposed until now, especially multi--qubit gates and quantum
gates on encoded qubits, can be implemented with our scheme. 4. We simulate
the projectors as quantum Hamiltonians and implement the quantum amplitude
amplification algorithm.

In conclusion we would like to stress that, in the simulations discussed in
this paper, the Hamiltonians involved are given only by conditional
displacement operators and conditional rotation operators. These operators
can be realized experimentally, e.g., in ion-traps. Therefore we believe that
the simulation strategies discussed in this paper have direct practical
relevance.

\acknowledgments
X. Wang acknowledge illuminating discussions with Klaus M\o lmer and Anders
S\o rensen on this subject. The authors thanks Dominik Janzing, Martin
R\"{o}tteler, Pawel Wocjan, Paolo Giorda, Mang Feng and Roman R. Zapatrin
for helpful discussions. This work has been supported by the European
Community through grant ISI-1999-10596 (Q-ACTA).


\begin{references}
\bibitem{Feynman}  R. P. Feynman, Int. J. Theor. Phys. {\bf 21}, 476 (1982)

\bibitem{Qc}  Special issue on quantum information, Phys. World {\bf 11}
33-57 (1998).

\bibitem{Lloyd1}  S. Lloyd, Science {\bf 273} 1072 (1996).

\bibitem{Control}  H. Rabitz, R. De Vivie-Riedle, M. Motzkus and K. Kompa,
Science {\bf 288}, 824 (2000); L. Viola, S. Lloyd, and E. Knill, Phys. Rev.
Lett.{\bf \ 83}, 4888 (1999); D. Vitali and P. Tombesi, Phys. Rev. A {\bf 59}%
, 4178 (1999); S. G. Schirmer, H. Fu and A. I. Solomon, Phys. Rev. A. 63,
063410 (2001); H. Fu, S. G. Schirmer, and A. I. Solomon, J. Phys. A: Math.
Gen. {\bf 34,} 1679 (2001).

\bibitem{QIP}M. A. Nielsen and I. L. Chuang, {\it Quantum computation and quantum information} (Cambridge University Press, Cambridge, 2000).

\bibitem{Marcus}  M. Stollsteimer and G. Mahler, Phys. Rev. A {\bf 64},
052301 (2001).

\bibitem{Dodd}  J. L. Dodd, M. A. Nielsen, M. J. Bremner, and R. T. Thew,
quant-ph/0106064; M. A. Nielsen, M. J. Bremner, and J. L. Dodd,
quant-ph/0109064.

\bibitem{Wocjan}  P. Wocjan, D. Janzing and Th. Beth, quant-ph/0106077; P.
Wocjan, M. R\"{o}tteler, D. Janzing, and Th. Beth, quant-ph/0109063;
quant-ph/0109088.

\bibitem{Bennett1}  C. H. Bennett, J. I. Cirac, M. S. Leifer, D. W. Leung,
N. Linden, S. Popescu, and G. Vidal, quant-ph/0107035; W. Leung,
quant-ph/0107041; G. Vidal and J. I. Cirac, quant-ph/0108076.

\bibitem{Chen}  H. Chen, quant-ph/0109115.

\bibitem{Deutsch}S. Lloyd, Phys. Rev. Lett. {\bf 75}, 346 (1995); D. Deutsch, A. Barenco, and A. Ekert, Proc. R. Soc. London A 
{\bf 449}, 669 (1995); D. P. DiVincenzo, Phys. Rev. A {\bf 51}, 1015 (1995).

\bibitem{Monroe96}  C. Monroe, D. M. Meekhof, B. E. King, and D. J.
Wineland, Science {\bf 272}, 1131 (1996).

\bibitem{Gerry}  C. C. Gerry, \pra {\bf 55}, 2478 (1997); C. C. Gerry and R. Grobe, {\bf 56}, 2390 (1997); {\bf 57}, 2247 (1998).

\bibitem{Milburn1}  G. J. Milburn, quant-ph/9908037.

\bibitem{Anders}  A. S\o rensen and K. M\o lmer, Phys. Rev. A {\bf 62},
022311 (2000).

\bibitem{Wangate}  X. Wang, A. S\o rensen and K. M\o lmer, Phys. Rev. Lett. 
{\bf 86}, 3907 (2001).

\bibitem{Hybrid}  S. Lloyd, quant-ph/0008057.

\bibitem{Gates}  D. P. Divincenzo, Proc. R. Soc. Lond. A {\bf 454}, 261
(1998); X. Zhou, D. W. Leung, I. L. Chuang, Phys. Rev. A {\bf 62,} 052316
(2000).

\bibitem{Shor}  P. W. Shor, SIAM J. Computing {\bf 26}, 1484 (1997).

\bibitem{Grover}  L. K. Grover, \prl  {\bf 79}, 325 (1997); {\bf 80}, 4329
(1998).

\bibitem{Phase}  S. Chaturvedi, M. S. Sriam and V. Srinivasan, J. Phys. A:
Math. Gen. {\bf 20}, L1071 (1987); R. G. Littlejohn, Phys. Rev. Lett. {\bf 61%
}, 2159 (1988); R. Simon and N. Kumar, J. Phys. A: Math. Gen. {\bf 21}, 1725
(1988); G. S. Agarwal and R. Simon, Phys. Rev. A {\bf 42}, 6924 (1990); M.
Dima, quant-ph/9912045.

\bibitem{Luis}  A. Luis, J. Phys. A: Math. Gen. {\bf 34}, 7677 (2001).

\bibitem{Berry}  M. V. Berry, Proc. R. Soc. London Ser. A {\bf 392}, 45
(1984).

\bibitem{AA}  Y. Aharonov and J. Anandan, Phys. Rev. Lett. {\bf 58}, 1593
(1987).

\bibitem{Paolo}  P. Zanardi and M. Rasetti, Phys. Lett. A {\bf 264}, 94
(1999); J. Pachos and P. Zanardi, Int. J. Mod. Phys. B {\bf 15}, 1257
(2001); J. Pachos, P. Zanardi, and M. Rasetti, Phys. Rev. A {\bf 61}, 010305(R)
(2000).

\bibitem{Jones}  J. A. Jones, V. Vedral, A. Ekert, and G. Castagnoli, Nature 
{\bf 403}, 869 (2000).

\bibitem{Fazio} G. Falci, R. Fazio, G. M. Palma, J. Siewert, V. Vedral
Nature {\bf 407}, 355 (2000).

\bibitem{Luming}  L. M. Duan, J. I. Cirac, and P. Zoller, Science {\bf 292},
1695 (2001).

\bibitem{Xiang}  Xiang-bin Wang and M. Keiji, Phys. Rev. Lett. {\bf 87}, 097901
(2001).

\bibitem{GHZ}  D. M. Greenberger, M. A. Horne, and Z. Zeilinger, {\it Bell\
's \ Theorem,\ Quantum\ Theory\ and\ Conceptions\ of\ the\ Universe}, edited
by M. Kafatos (Kluwer Academic, Dordrecht, 1989), p. 69; D. Bouwmeester, A.
Ekert and A. Zeilinger (Eds.), {\it \ The Physics of Quantum Information}
(Springer-Verlag, Berlin, 2000).

\bibitem{Molmer}  K. M\o lmer and A. S\o rensen, \prl {\bf 82}, 1835 (1999).

\bibitem{Ueda}  M. Kitagawa and M. Ueda, Phys. Rev. A {\bf 47}, 5138 (1993).

\bibitem{Three}  C. H. Tseng, S. Somaroo, Y. Sharf, E. Knill, 
R. Laflamme, T. F. Havel, and D. G. Cory, \pra {\bf 61}, 012302 (2000).

\bibitem{Cnot}A. Barenco, C. H. Bennett, R. Cleve, D. P. DiVincenzo, N. Margolus, 
P. Shor, T. Sleator, J. A. Smolin, and H.Weinfurter, \pra {\bf 52}, 3457 (1995); 
J. I. Cirac and P. Zoller, \prl {\bf 74}, 4091 (1995).

\bibitem{Heis1}  W. Jones and N. March, {\it Theoretical Solid State Physics}
(Dover, New York, 1985), Vol. 1.

\bibitem{Heismulti}  N. Sourlas, Nature {\bf 339}, 693 (1989); I. Kanter and D.
Saad, \prl  {\bf 83}, 2660 (1999).

\bibitem{KimLee}  J. Kim, J. S. Lee and S. Lee, Phys. Rev. A {\bf 61}, 032312
(2000).

\bibitem{CP}  D. P. DiVicenzo, Science {\bf 270}, 255 (1995); E. Solano, M.
Franca Santos, and P. Milman, Phys. Rev. A {\bf 64,} 024304 (2001).

\bibitem{Fredkin}  E. Fredkin and T. Toffoli, Int. J. Theor. Phys. {\bf 21},
219 (1982).

\bibitem{Zurek}  W. H. Zurek and R. Laflamme, quant-ph/9605013.

\bibitem{Steane}  A. M. Steane, \prl  {\bf 77}, 793 (1996).

\bibitem{DFS}  P. Zanardi and M. Rasetti, Mod. Phys. Lett. B {\bf 11}, 1085
(1997); \prl  {\bf 79}, 3306 (1997); L. M. Duan and G. C. Guo, Phys. Rev.
Lett. {\bf 79}, 1953 (1997); D. A. Lidar, I. L. Chuang, and K. B. Whaley, %
\prl  {\bf 81}, 2594 (1998); D. A. Lidar, D. Bacon, and K. B. Whaley, 
\prl {\bf 82}, 4556 (1999).

\bibitem{Bennett}  C. H. Bennett, D. P. DiVincenzo, J. A. Smolin, and W. K.
Wootters, \pra {\bf 54}, 3824 (1996); R. Laflamme, C. Miquel, J. P. Paz, and
W. H. Zurek, \prl  {\bf 77}, 198 (1996).

\bibitem{Hwang}  W. Y. Hwang, H. Lee, D. D. Ahn, and S. W. Hwang, Phys. Rev. A 
{\bf 62}, 062305 (2000)

\bibitem{NMRgrover}  I. L. Chuang et al., Phys. Rev. Lett. {\bf 80}, 3408
(1998); J. A. Jones, M. Mosca, R. H. Hansen, Nature (London) {\bf 393}, 344
(1998).

\bibitem{Hoyer}  G. Brassard, P. H\o yer, and A. Tapp. {\em Proc. of 25th
International Colloquium on Automata, Languages, and Programming (ICALP'98)}%
, Vol. 1443 of Lecture Notes in Computer Science, pp. 820-831, 1998; P. H\o
yer, Phys. Rev. A {\bf 62}, 052304 (2000).

\bibitem{Long}  G. L. Long, Phys. Rev. A, {\bf 64}, 022307 (2001); G. L.
Long, L. Xiao, and Y. Sun, quant-ph/0107013.

\bibitem{Lidafa}  D. Li and X. Li, Phys. Lett. A {\bf 287}, 304 (2001).
\end{references}
\end{document}